\documentclass[aps,prl,twocolumn,floats]{revtex4}
\usepackage{epsfig}
\usepackage{epsf}
\usepackage{hyperref}
\usepackage{graphicx}
\usepackage{latexsym}
\usepackage{rotating}
\begin{document}

\bibliographystyle{unsrt}

\title{Excitonic effects and optical properties of passivated CdSe clusters}

\author{Marie Lopez del Puerto$^{(1)}$, Murilo L. Tiago$^{(2)}$
, and James R. Chelikowsky$^{(1,2,3)}$}
\affiliation{
 $^{(1)}$ Department of Physics, 
 University of Minnesota, 
Minneapolis, MN 55455.\\
$^{(2)}$ Center for Computational Materials, Institute for
 Computational Engineering and Sciences, University of Texas, Austin, TX 78712.\\
 $^{(3)}$ Departments of Physics and Chemical Engineering, University of Texas, Austin, TX 78712.}
\date{\today}

\begin{abstract}
We calculate the optical properties of a series of passivated
non-stoichiometric CdSe clusters using two first-principles approaches: time-dependent
density functional theory within the local density
approximation, and by solving the Bethe-Salpeter equation for optical
excitation with the GW approximation for the self-energy.
We analyze the character of optical excitations leading to
the first low-energy peak in the absorption cross-section of these
clusters. Within time-dependent density functional theory, we find
that the lowest-energy excitation is mostly a single-level to
single-level transition. In contrast, many-body methods predict
a strong mixture of several different transitions, which is a signature of exciton effects. We also find that the majority of
the clusters have a series of dark transitions before the first bright
transition. This may explain the long radiative lifetimes observed experimentally. 

\end{abstract}

\maketitle

Experimental advances in the synthesis of semiconductor clusters have
 stimulated considerable theoretical effort to understand the
 optical and electronic properties of these systems.
Semiconductor clusters often exhibit strong
size-dependent effects, which are not yet fully
understood. As an intermediate system between single atoms and bulk
materials, semiconductor clusters are also of intrinsic theoretical
interest. Clusters of II-VI elements, such as CdSe, have attracted considerable
attention in recent years owing to 
their potential technological applications in various devices such as solar cells,
 lasers and biological imaging tools, among others
\cite{applications}. Great effort has gone into fabricating
and characterizing size-controlled samples. This is particularly
challenging because clusters of different sizes have similar
 stoichiometry and are synthetized by
 similar reactions in which the temperature, the
 solvent and the ratio and concentration of precursors have to be
 carefully controlled \cite{soloviev}. Theoretical calculations are
 difficult because of the inherent complexity of accurate theories, as
 we can infer from the limited number of theoretical articles in the
 literature \cite{troparevsky,vandriel,eichkorn,tiago}.

The use of pseudopotentials and density functional theory (DFT) have been
very successful in determining the ground-state properties of both
bare \cite{troparevsky} and passivated \cite{eichkorn}  CdSe
clusters. However, DFT is a ground-state theory, and it has serious
shortcommings in providing a quantitative description of optical
and electronic excitations \cite{onida,rohlfing}.
Understanding spectroscopic experiments requires
the computation of excited state properties, which present a greater
 challenge than ground-state calculations.  The problem is addressed by both time-dependent
 density functional theory (TDDFT) and GW/Bethe-Salpeter
 (GW/BSE) methods. TDDFT is simpler to implement, but
 a good general approximation for the exchange-correlation functional is still
 lacking \cite{onida}. The local-density approximation within TDDFT
 (TDLDA) has been
 found to give accurate results for some finite systems such
 as sodium clusters
 \cite{vasiliev_2}. In others, it gives at best a qualitative picture \cite{onida,tiago,delpuerto}.  On the other hand,
 GW/BSE has been shown to be very accurate in bulk materials \cite{onida,rohlfing}, albeit more
 computationally demanding. Until recently, first-principles GW/BSE
 calculations have been done only for very small clusters, containing
 no more than 35 atoms \cite{onida,rohlfing,benedict,tiago}.

In this letter we present a detailed comparison of TDLDA and GW/BSE
calculations of the optical properties of passivated CdSe
clusters. By examining the character of the transition leading to the
first peak in the absorption spectra, we investigate the importance of
many-body effects, fully accounted for in GW/BSE but absent in TDLDA.
Calculations of energy band gaps and absorption
 spectra of bare CdSe clusters
 have been done using TDDFT within the local-density approximation
 (TDLDA) \cite{troparevsky}. Energy band gaps of passivated CdSe
 clusters have also been calculated within the TDDFT framework
 \cite{eichkorn}. GW/BSE has been used for calculations of optical
 properties of molecular
 systems, silicon clusters, and in crystals \cite{onida,rohlfing,
   benedict, behrens}, but not III-V or II-VI
 clusters yet. Experimental data on CdSe clusters is
 readily available \cite{soloviev} which makes these systems ideal
 for a comparative analysis.

\begin{figure}
\begin{centering}
\includegraphics[width=0.6in]{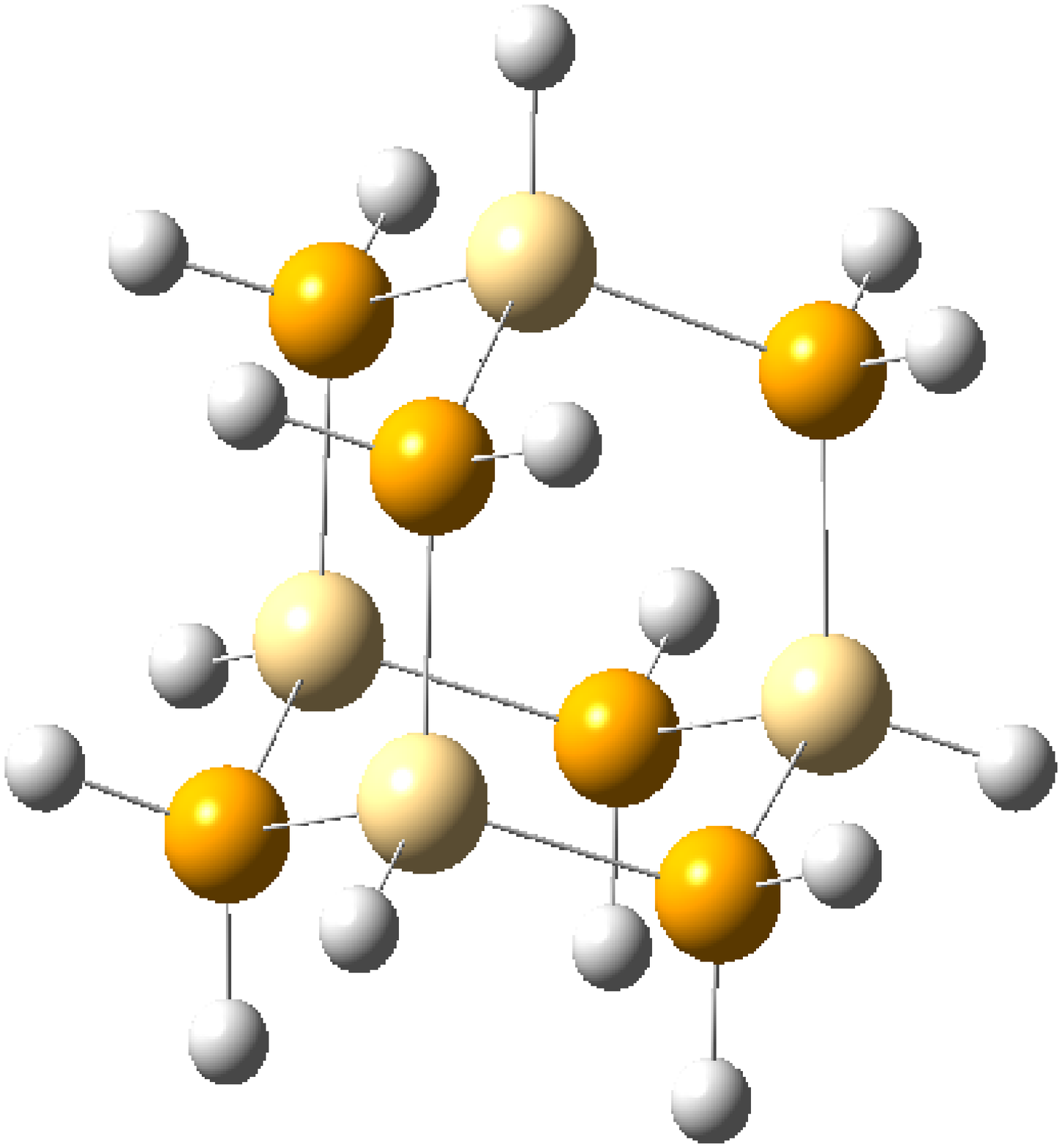}%
\hspace{15pt}%
\includegraphics[width=0.75in]{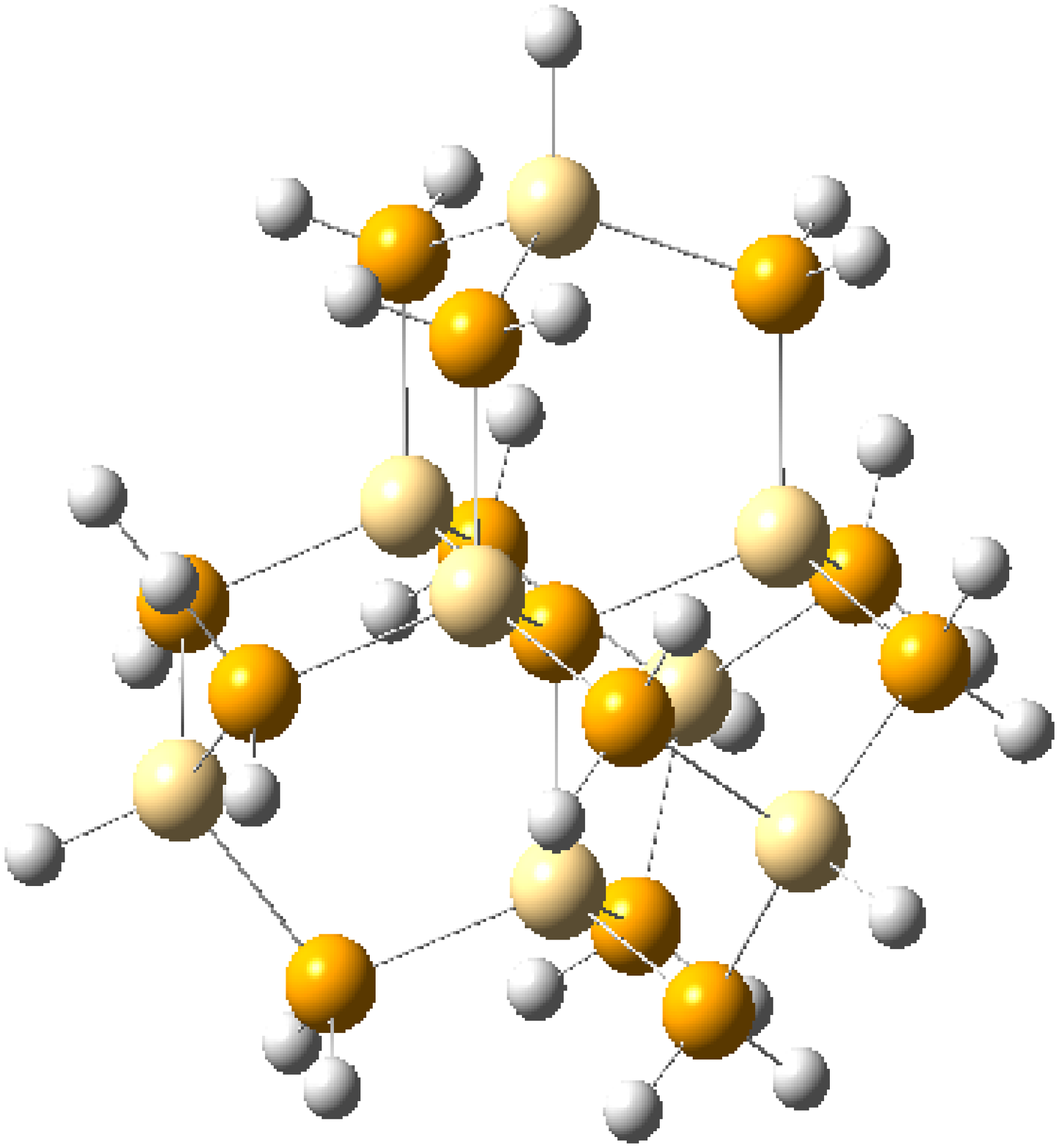}%
\hspace{15pt}
\includegraphics[width=0.825in]{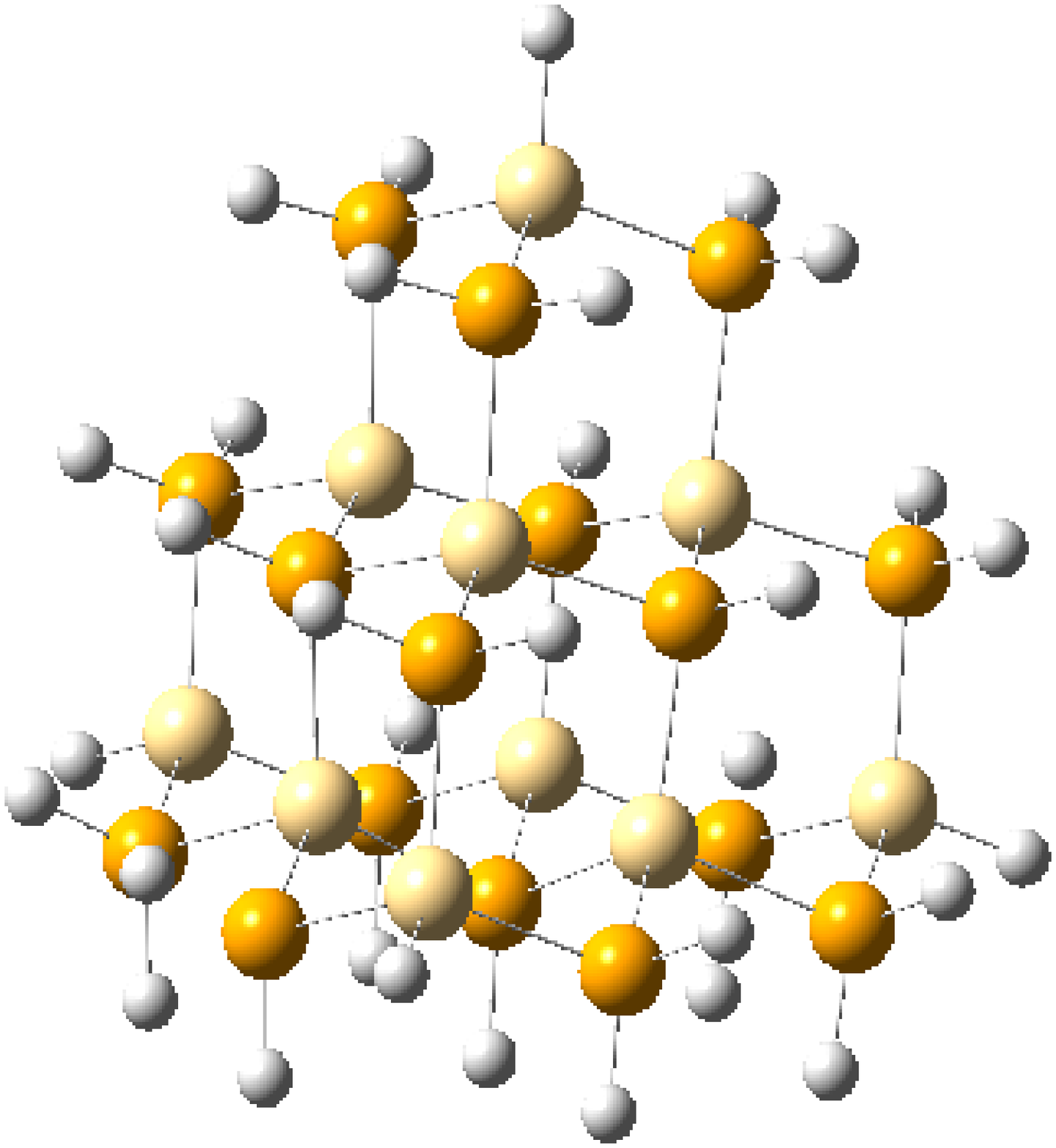}\\
\includegraphics[width=0.9in]{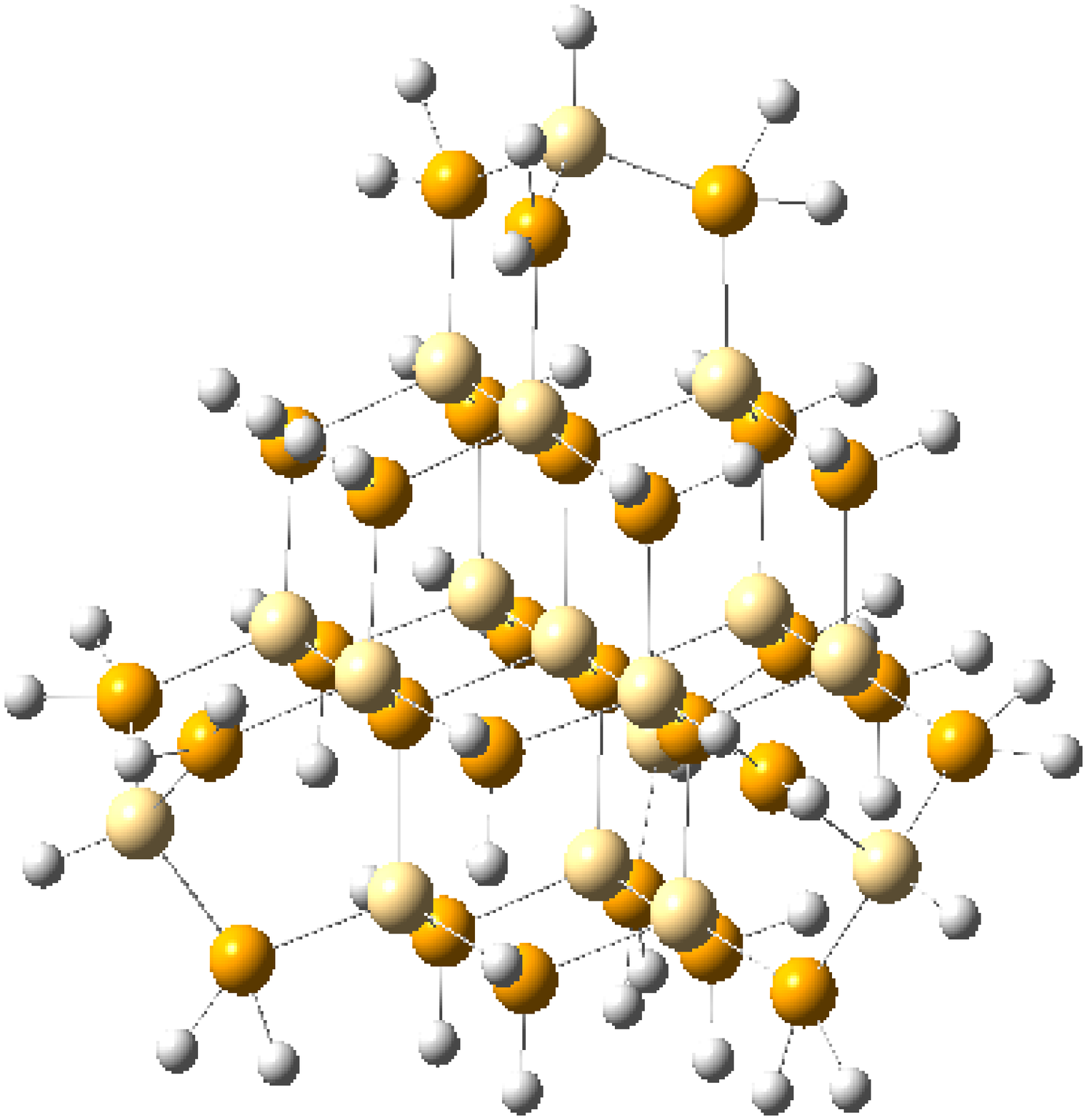}%
\hspace{15pt}%
\includegraphics[width=1.05in]{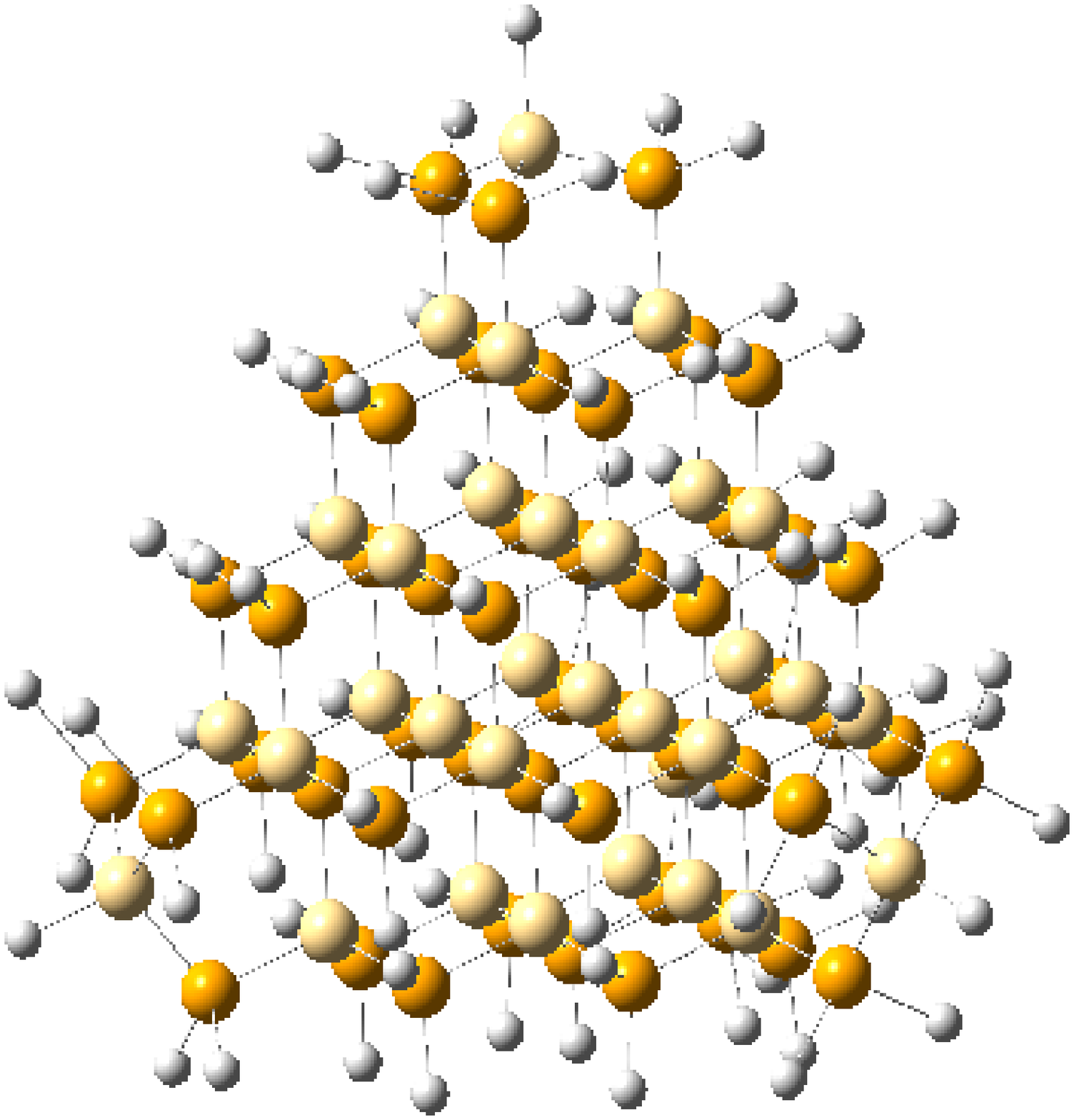}
\caption{Geometry of the clusters studied: Cd$_4$Se$_6$,
  Cd$_8$Se$_{13}$, Cd$_{10}$Se$_{16}$, Cd$_{17}$Se$_{28}$ and
  Cd$_{32}$Se$_{50}$. Cd atoms are dark yellow, Se atoms are light yellow, and fictitious H atoms are small gray. }
\label{fig1}
\end{centering}
\end{figure}

We studied a series of five clusters: Cd$_4$Se$_6$, Cd$_8$Se$_{13}$,
Cd$_{10}$Se$_{16}$, Cd$_{17}$Se$_{28}$ and Cd$_{32}$Se$_{50}$. While
larger CdSe clusters are found to be spherical \cite{katari},
these smaller clusters are of pyramidal shape \cite{soloviev} (see Fig. \ref{fig1}). Two of
the clusters studied have zincblende structures
(Cd$_4$Se$_6$ and Cd$_{10}$Se$_{16}$), while the rest are of the
wurtzite type. All clusters were passivated by
 fictitious hydrogen atoms of charge 1.5$e$ (attached to surface Cd atoms) and
 0.5$e$ (attached to surface Se atoms) \cite{huang}, in order to
 simulate the effect of surfactants on the surface of the real
 clusters \cite{soloviev}.

 The initial geometry of the clusters was constructed  based on the
 X-ray data \cite{soloviev} and then relaxed \cite{byrd, zhu}. Tetrahedral symmetry was conserved during the relaxation. The Cd-Se bonds in the final structures are comparable to the experimental measurements \cite{soloviev}, but about 5\% shorter than the Cd-Se bond in bulk cadmium selenide.  Calculations were done using 
norm-conserving pseudopotentials \cite{troullier} constructed within
the local density approximation of density functional
theory. Pseudopotentials for cadmium and selenium have scalar
 relativistic effects included and the interaction between the Cd $4d$
 orbital and valence orbitals is accounted for by a non-linear core
 correction in the pseudopotential \cite{louie}. Spin-orbit and
 semicore effects beyond the non-linear core correction are
 ignored. In order to estimate how important these effects are, we
 computed the band gap of bulk CdSe in the wurtzite structure. The gap
 obtained within the GW approximation is 1.8 eV, which compares well
 with the spin-orbit averaged gap obtained in experiment: 1.97 eV
 \cite{zakharov}. The measured spin-orbit splitting, $\Delta$, at the valence band maximum
 at the $\Gamma$ point is $\Delta = 0.43 eV$
 \cite{zakharov}. Including spin-orbit splitting would change the
 calculated energy gap by $ E_{gap} = \frac{\Delta}{3} \sim 0.15 eV $.   The Kohn-Sham equations are solved on a real
 space grid using
a higher-order finite difference method \cite{chelikowsky}.  All calculations were
done within a spherical boundary of radius at least 6 a.u. from the
outermost passivating atoms. A grid spacing of 0.3 a.u. was used for
 LDA calculations, while a grid spacing of 0.6 a.u. was used for the
 calculation of optical properties as described below.

In TDLDA \cite{casida, vasiliev}, the optical response is evaluated as a
first-order perturbation in the electron density due to an external
potential. The excitation energies $\Omega_n$ are obtained from a
solution of the eigenvalue equation \cite{casida, vasiliev}:

\begin{equation}
{\bf QF}_n=\Omega_n^2{\bf F}_n
\label{omegan}
\end{equation}

\noindent The matrix elements for ${\bf Q}$  are given by
\begin{equation}
Q_{ij\sigma,kl\tau} = \delta_{i,k}\delta_{j,l}\delta_{\sigma,\tau}\hbar^2\omega^2_{kl\tau}+2\hbar\sqrt{\lambda_{ij\sigma}\omega_{ij\sigma}}K_{ij\sigma}^{kl\tau}\sqrt{\lambda_{kl\tau}\omega_{kl\tau}},
\end{equation}

\noindent
where $\lambda_{kl\tau}=n_{l\tau}-n_{k\tau}$ are the difference
between the occupation numbers, and $\hbar
\omega_{lk\tau}=\epsilon_{k\tau}-\epsilon_{l\tau}$ is the difference
between the eigenvalues of the single-particle states. $K$ is the
coupling matrix which describes the linear response of the system \cite{casida}.

Electron-phonon
coupling and temperature dependence effects are included {\it a
  posteriori} by broadening the absorption spectra with a
normalized Gaussian function with fixed dispersion of $0.1 \; eV$.

In the GW/BSE method, the many-body expression for the polarizability
$\Pi$ is related to the electron-hole correlation function $L$ by:

\begin{equation}
\Pi (1,2) = - i L(1,2;1^+,2^+)
\end{equation}

\noindent where $L$ satisfies the Bethe-Salpeter equation \cite{onida, rohlfing}:

\begin{eqnarray}
L(1,2;3,4) = G(1,4)G(2,3) + \nonumber \\
\int d(5678)G(1,5)G(6,3)K(5,7;6,8)L(8,2;7,4).
\label{eq2}
\end{eqnarray}

Solving the above equation, we obtain optical excitations of the
electronic system. $G$ is the electron Green's function,
and the kernel operator $K$ describes interactions between the excited
electron and the hole left behind in the electron sea. The electron self-energy is calculated within the GW approximation
\cite{hybertsen}. As in TDLDA, the absorption spectra is broadened by
normalized Gaussian functions with fixed dispersion. In both TDLDA and
GW/BSE, the optical gap is defined as the energy of the first
transition with measurable oscillatory strength (the first allowed
transition) \cite{vasiliev}.

\begin{figure}
\begin{centering}
\includegraphics[width=3.in]{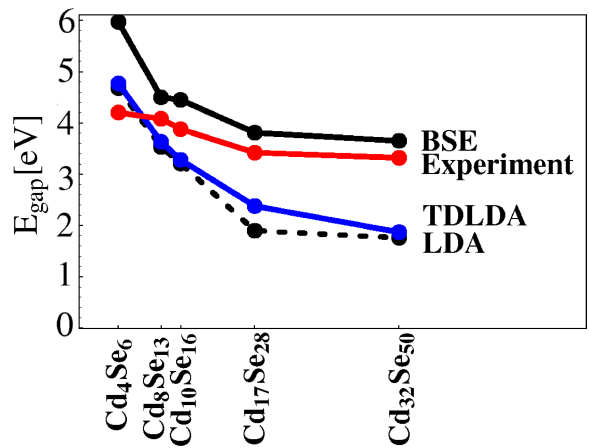}
\caption{Experimental and calculated optical gaps. The LDA gap is
  simply the difference between Kohn-Sham eigenvalues.}
\label{fig2}
\end{centering}
\end{figure}

Figure \ref{fig2} shows optical gaps as
a function of CdSe cluster size. For all but the smallest cluster,
our calculations
\cite{eichkorn} show a trend very similar to that found in
experiment. The dependence of the gap on cluster size is
very strong. It is interesting to note that while TDLDA calculations
underestimate the gap by $\sim$0.5 eV to $\sim$1.5 eV, GW/BSE
overestimates the gap by less than 0.6 eV. The discrepancy between
experimental data and TDLDA calculations increases as the size of the
cluster increases, but the opposite is observed for GW/BSE
calculations. Based on the analysis made for bulk CdSe, we expect that
neglected spin-orbit and semicore effects at pseudopotential level are
responsible for a residual discrepancy between theory and experiment
of a few tenths of electron-volt.

There is a large discrepancy between the theoretical calculations and the experimental measurements for the optical gap of the smallest cluster, Cd$_4$Se$_6$. For the larger clusters, all TDLDA results are below experiment, while BSE results are above. This is not the case for Cd$_4$Se$_6$, as both TDLDA and BSE predict larger optical gaps than measured experimentally. Given the reduced number of atoms in this cluster, it is not clear whether the organic ligands on the surface or some other mechanism might be responsible for this discrepancy.

\begin{figure}
\begin{centering}
\includegraphics[width=3.2in]{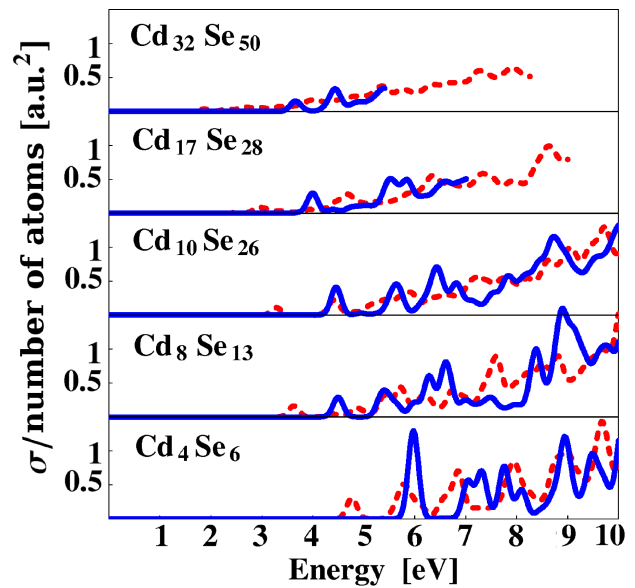}
\caption{Absorption cross section calculated within TDLDA (dashed curve) and
  GW/BSE (solid curve).}
\label{fig3}
\end{centering}
\end{figure}

Calculated absorption cross-sections (normalized by the total number of cadmium and selenium atoms in each cluster) are shown in
Fig. \ref{fig3}.  Both TDLDA and BSE have a well-defined first peak at
low energy, with a second peak separated from the first by $\sim$1
eV. We have analyzed the character of the excitation leading to the
first peak observed in the absorption cross-section. As shown in Table
\ref{table2}, the first TDLDA excitation in all clusters is mostly a result of single-level to single-level transitions. For
the three smallest clusters, the
dominant transition is from the HOMO (highest occupied molecular orbital, which is triple-degenerate for most
clusters, without spin-orbit splitting) to the LUMO (lowest unoccupied molecular orbital, which is non-degenerate in all cases). This
is not so  for the two largest clusters  as
they have a series of dark transitions (transitions with negligible oscillatory
strength) before the first bright (allowed) transition. The first
optically allowed transition for Cd$_{17}$Se$_{28}$ involves
transitions from the third level below the HOMO to the LUMO. For Cd$_{32}$Se$_{50}$ the first allowed transition is from the
first level below the HOMO to the LUMO.

\begin{table}[b]
\centering
\caption{Energy of the first allowed transition, E; presence of dark transitions
  (those with negligible oscillatory strength) before the first allowed transition; and percentage of
  the lowest energy peak in the absorption cross section that is due
  to the single-level to single-level transition indicated in the text. No entry in the percentage column indicates that the
  transition is strongly mixed, i.e, the largest component contributes less than 25\%. }
\label{table2}
\begin{tabular}{l|ccc|ccc}
\hline
& \multicolumn{3}{c|}{\bf TDLDA} & \multicolumn{3}{c}{\bf GW/BSE} \\
Cluster &  E [eV] \hspace{5pt} & Dark \hspace{5pt} & \%  & E
[eV]\hspace{5pt}  & Dark \hspace{5pt} & \% \\
\hline
Cd$_4$Se$_6$ & 4.77 & no & 96 &  5.97 &  yes & 91\\
\hline
Cd$_8$Se$_{13}$ & 3.53 & no & 94 & 4.50 & no & \\
\hline
 Cd$_{10}$Se$_{16}$ & 3.21 & no & 96 & 4.45 & yes & \\
\hline
 Cd$_{17}$Se$_{28}$ &  2.38 & yes & 98 & 3.81 & yes & \\
\hline
 Cd$_{32}$Se$_{50}$ &  1.87 & yes & 98 & 3.65  & yes & \\
\hline
\end{tabular}
\end{table}

Experimentally, bulk CdSe in the wurtzite and zincblende structures is
a direct gap semiconductor \cite{edwards}. Our own LDA calculations
agree with these experimental results. Clusters, however, do not
always behave in the same way. This can be attributed to confinement
effects and the geometry of each cluster \cite{tolbert, krishna}.
Cd$_4$Se$_6$, Cd$_8$Se$_{13}$ and Cd$_{10}$Se$_{16}$ have dipole
allowed transitions between the triple-degenerate HOMO and the
non-degenerate LUMO. But the HOMO of Cd$_{17}$Se$_{28}$ is
non-degenerate because of a change in the ordering of the energy
levels (the HOMO-1 is triple-degenerate), and the HOMO-LUMO transition
becomes dipole forbidden
as a result of selection rules. Cd$_{32}$Se$_{50}$ has a dipole
allowed HOMO-LUMO transition, but the oscillator strength is small
because of little overlap between HOMO and LUMO wavefunctions.

The character of the
GW/BSE excitation leading to the first peak in the absorption spectra
is very different from the one predicted by TDLDA. For the
smallest cluster the excitation is still dominated by a single-level to
single-level (HOMO $\rightarrow$ LUMO) transition. For the rest of the clusters, however, the
excitation is the result of a strong mixture of different
transitions.  There are two sources of mixing in GW/BSE: at the GW
level, mixing occurs because of fact that LDA wavefunctions are not
identical to quasiparticle wavefunctions \cite{hybertsen,rohlfing,tiago};
at the BSE level, the electron-hole kernel is stronger and more
non-local than the TDLDA kernel. We find that the mixing in TDLDA is
one order of magnitude smaller than the mixing in GW/BSE. This is a
result of stronger coupling matrix elements at the GW/BSE level
compared to TDLDA. We quantify the mixing as

\begin{equation}
M = \frac{\Sigma_i \Sigma_{j \neq i} |H_{ij}|^2}{\Sigma_i |H_{ii}|^2}
\end{equation}

\noindent where $H$ is the effective Hamiltonian matrix in either
TDLDA or GW/BSE methods. $M=0$ corresponds to a situation where there
is no mixing between different single electron transitions. We find
that the mixing $M$ within GW/BSE is of the order of $10^{-2}$ while
within TDLDA it is of the order of $10^{-3}$. In both cases the mixing
is non-zero but still
much smaller than one. As a result of the stronger mixing in GW/BSE, the
excitation cannot be associated with a single
electron-hole transition. A similar behavior has been observed in
small silicon clusters and in bulk semiconductors
\cite{onida,rohlfing}, and it is a signature of excitonic effects.

The effects of mixing also explain the observed divergence in the
experimental and TDLDA curves in Fig. \ref{fig2}.  Since
the energy levels in the smaller
clusters are more separated, the mixing effect is not as large and TDLDA calculations are
more accurate than for larger clusters where the energy levels are
closer together and more mixing can occur.

In the GW/BSE calculations we find that there are dark transitions for all clusters but
Cd$_8$Se$_{13}$. Van Driel {\it et al.} \cite{vandriel} have recently 
shown that measured rates of emission are completely determined by radiative decay and that the occupation of dark excitonic states considerably attenuates spontaneous emission. The presence of dark transitions in our GW/BSE calculations then may explain in part
  the long radiative lifetimes ($\sim$1-10$\mu$s) observed experimentally.

In conclusion, we have calculated the optical properties of a series
of small CdSe clusters using two different approaches: TDLDA and
GW/BSE. We find that the two methods lead to a very different
character for the lowest energy excitation. In TDLDA, the excitation
is dominated by a single-level to single-level transition. In GW/BSE,
however, the excitation is the result of strong mixing between
different transitions. We interpret this as due to exciton effects. Our
calculations also show that most clusters have a series of dark
transitions before the first bright transition, which attenuates spontaneous emission and may explain the long radiative lifetimes of these clusters.

We would like to acknowledge helpful discussions with Xiangyang Huang. This work was supported in part by the
National Science Foundation under DMR-0551195 and the U.S. Department of Energy under DE-FG02-89ER45391 and DE-FG02-03ER15491. Calculations were performed  at the Minnesota
Supercomputing Institute, the Texas Advanced Computing Center and at
the National Energy Research Scientific Computing Center (NERSC).

\end{document}